\documentclass[]{myspie}  

\usepackage[]{graphicx}

\title{The Cosmic Gamma-Ray Bursts and Their Host Galaxies in a Cosmological Context} 

\author{
S. G. Djorgovski\supit{a},
S. R. Kulkarni\supit{a},
D. A. Frail\supit{b},
F. A. Harrison\supit{a},
J. S. Bloom\supit{a},
E. Berger\supit{a},
P.~A.~Price\supit{a},
D. Fox\supit{a},
A. M. Soderberg\supit{a},
T. J. Galama\supit{a},
D. E. Reichart\supit{a},
R. Sari\supit{a},
S. Yost\supit{a},
A.~A.~Mahabal\supit{a},
S.~M. Castro\supit{a},
R. Goodrich\supit{c}, and
F. Chaffee\supit{c}
\skiplinehalf
\supit{a} Division of Physics, Mathematics, and Astronomy, Caltech, Pasadena, CA 91125, USA\\
\supit{b} National Radio Astronomy Observatory, Socorro, NM 87801, USA\\
\supit{c} W.M. Keck Observatory, Kamuela, HI 96743, USA
}

\authorinfo{\bf To appear in: Discoveries and Research Prospects from 6-10m Class
Telescopes, ed. P. Guhathakurta, Proc. SPIE, vol. 4834 (2003)}

\begin{document} 
\maketitle 


\begin{abstract}
Studies of the cosmic gamma-ray bursts (GRBs) and their host galaxies are now
starting to provide interesting or even unique new insights in observational
cosmology.  Observed GRB host galaxies have a median magnitude $R \sim 25$ mag,
and show a range of luminosities, morphologies, and star formation rates,
with a median redshift $z \sim 1.0$.
They represent a new way of identifying a population of star-forming galaxies
at cosmological redshifts, which is mostly independent of the traditional
selection methods.  They seem to be broadly similar to the normal field
galaxy populations at comparable redshifts and magnitudes, and indicate
at most a mild luminosity evolution over the redshift range they probe.
Studies of GRB optical afterglows seen in absorption provide a powerful
new probe of the ISM in dense, central regions of their host galaxies,
which is complementary to the traditional studies using QSO absorption
line systems.  Some GRB hosts are heavily obscured, and provide a new
way to select a population of cosmological sub-mm sources.
A census of detected optical tranistents may provide an important new
way to constrain the total obscured fraction of star formation over the
history of the universe.
Finally, detection of GRB afterglows at high redshifts ($z > 6$) may
provide a unique way to probe the primordial star formation, massive IMF,
early IGM, and chemical enrichment at the end of the cosmic reionization era.
\end{abstract}




\section{Introduction}

Ever since their discovery (Klebesadel, Strong, \& Olson 1973), the cosmic
gamma-ray bursts (GRBs) represented one of the outstanding puzzles of science.
A considerable progress in their understanding has been made since the
discovery of GRB afterglows in x-rays (Costa et al. 1997), visible
(van Paradijs et al. 1997), and radio (Frail et al. 1997), and the conclusive
establishment of their cosmological nature (Metzger et al. 1997).  Large telescopes
such as the Kecks and the VLT have played crucial roles in the cracking of this
cosmic mystery.  Today, studies of GRBs are one of the most active and vibrant
areas of astrophysics.

In this review we focus on some of the cosmological aspects of GRBs and their
host galaxies.  We do not address the issues of GRB afterglows and their
physics, beaming, and progenitor mechanisms, including the GRB-Supernova
connection.  Instead, we direct the reader to some other, recent reviews
which cover these subjects, e.g., Piran (1999), M\'esz\'aros (2001, 2002),
van Paradijs, Kouveliotou, \& Wijers (2000), Kulkarni et al. (2000),
Pian (2002), and many others.  Several recent conference volumes provide
many additional reviews and relevant papers.
Parts of the present text have also appeared in the review by Hurley, Sari,
\& Djorgovski (2003).

\section{GRB Host Galaxies and Redshifts}

Host galaxies of GRBs serve a dual purpose:  they determine the redshifts,
which are necessary for a complete physical modeling of the bursts,
and they provide some insights about the possible nature of the progenitors,
e.g., their relation to massive star formation, etc.  The subject has been
reviewed previously, e.g., by Djorgovski et al. (2001b, 2002), Hurley, Sari \&
Djorgovski (2003), and many others.

\subsection{Overall Properties of GRB Hosts}

As of the late 2002, plausible or certain host galaxies have
been found for all but 1 or 2 of the bursts with optical, radio, or x-ray
afterglows localised with arcsecond precision.  Two examples are shown in
Fig. 1.
The median apparent
magnitude is $R \approx 25$ mag, with tentative detections or upper limits
reaching down to $R \approx 29$ mag (Fig. 2).  The few missing cases are at least
qualitatively consistent with being in the faint tail of the observed
distribution of host galaxy magnitudes.

\begin{figure}
\begin{center}
\begin{tabular}{c}
\includegraphics[height=7cm]{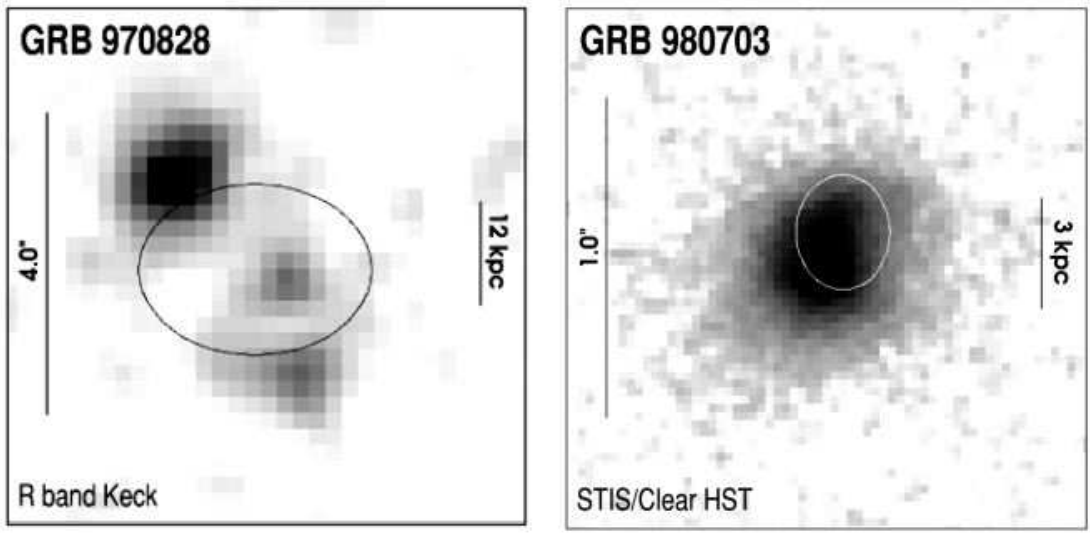}
\end{tabular}
\end{center}
\caption[example] { \label{fig:hostgals} 
Examples of GRB host galaxies, from the compilation by Bloom, Kulkarni, \& Djorgovski
(2002), based on the ground-based and HST imaging.  The ellipses indicate the 
positions of the radio transient (for GRB 970828) and optical transient (for
GRB 980703).  Vertical bars give the observed image scale in arcsec, and the
restframe scale in kpc.
}
\end{figure} 

Down to $R \sim 25$ mag, the observed distribution is consistent with deep
field galaxy counts (Brunner, Connolly, \& Szalay 1999),
but fainter than that, complex selection effects may be playing a role.
It can also be argued that the observed distribution should correspond
roughly to luminosity-weighted field galaxy counts.  However, the actual
distribution would depend on many observational selection and physical
(galaxy evolution) effects, and a full interpretation of the observed
distribution of GRB host galaxy magnitudes requires a careful modeling.
We note also that the observations in the visible probe the
UV in the restframe, and are thus especially susceptible to extinction. 
However, sub-mm detections of dusty GRB hosts are currently limited by the
available technology to only a handful of ultraluminous sources.

\begin{figure}
\begin{center}
\begin{tabular}{c}
\bigskip\bigskip
\includegraphics[height=7cm]{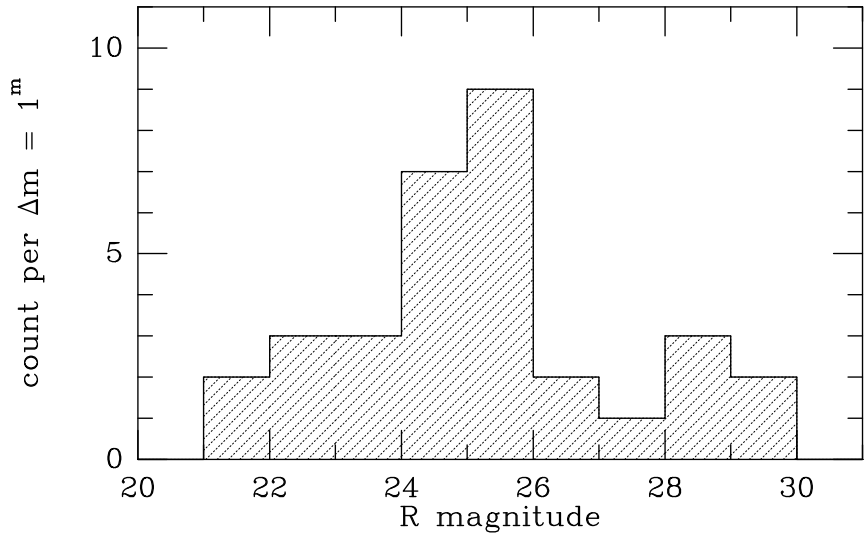}
\end{tabular}
\end{center}
\caption[example] { \label{fig:maghist} 
The distribution of observed GRB host galaxy magnitudes in the $R$ band, 
corrected for the Galactic extinction.  The bright end mimics the overall
shape of field galaxy counts, but the decline at $R > 25$ mag is $not$
due to the incompleteness: to date, only a couple of well-localised GRBs
lack convincing host detections.
}
\end{figure} 

Starting with the first redshift measurement which unambiguosly demonstrated
the cosmological nature of GRBs (Metzger et al. 1997)
there are now (late 2002) over 30 redshifts measured for GRB hosts 
and/or afterglows.  The median redshift is $\langle z \rangle \approx 1.0$,
spanning the range from 0.25 (or 0.0085, if the association of GRB 980425
with SN 1998bw is correct)
to 4.5 (for GRB 000131).
The majority of redshifts so far are from the spectroscopy of host galaxies,
but an increasing number are based on the absorption-line systems seen in
the spectra of the
afterglows (which are otherwise featureless power-law continua).  
Figure 3 
shows two examples.  Reassuring
overlap exists in several cases; invariably, the highest-$z$ absorption system
corresponds to that of the host galaxy, and has the strongest lines.
In some cases (a subset of the so-called ``dark bursts'') no optical
transient (OT) is detected, but a combination of the X-ray (XT) and radio
transient (RT) unambiguously pinpoints the host galaxy.

\begin{figure}
\begin{center}
\begin{tabular}{c}
\includegraphics[height=7cm]{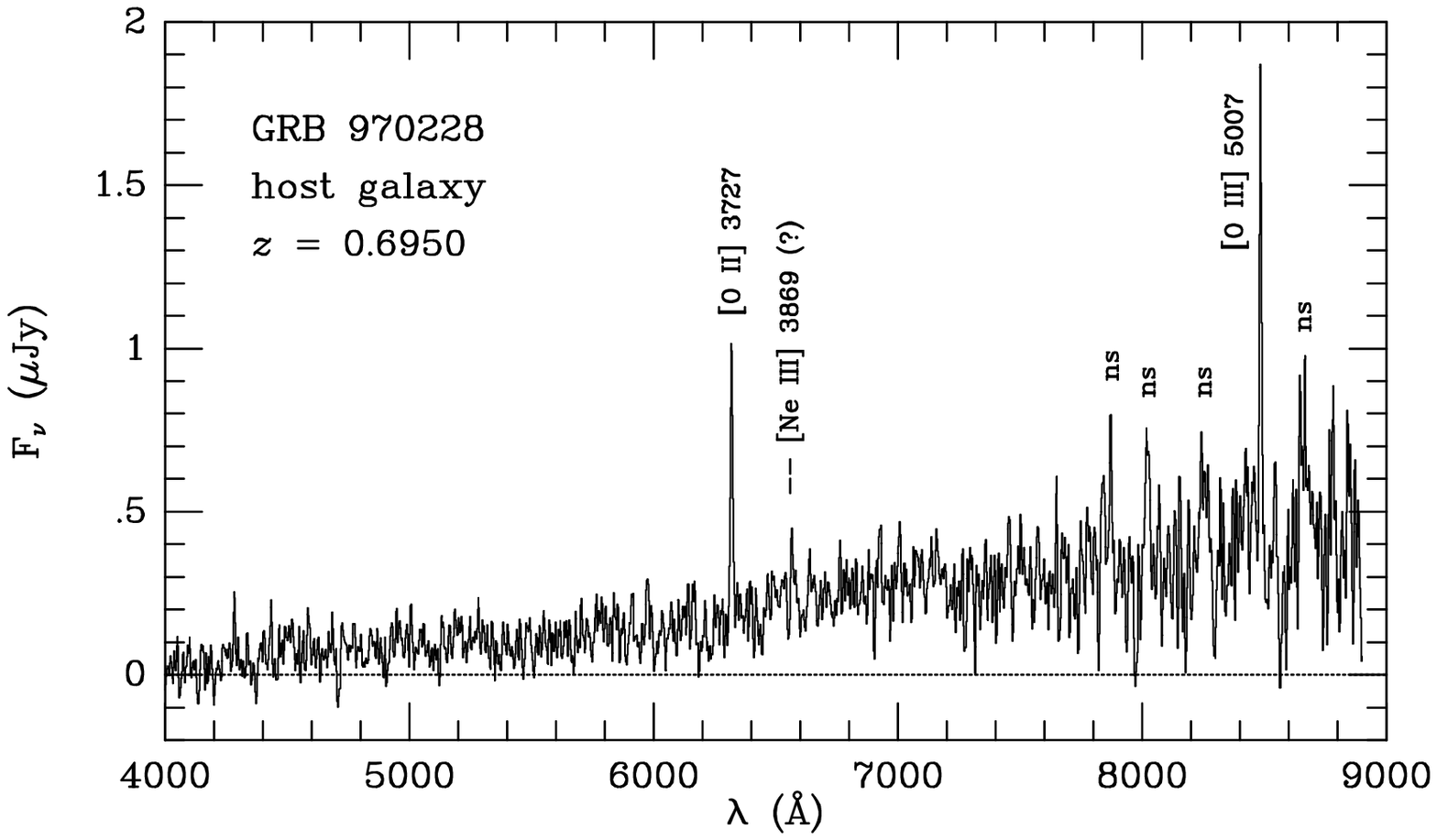}
\end{tabular}
\end{center}

\begin{center}
\begin{tabular}{c}
\includegraphics[height=7cm]{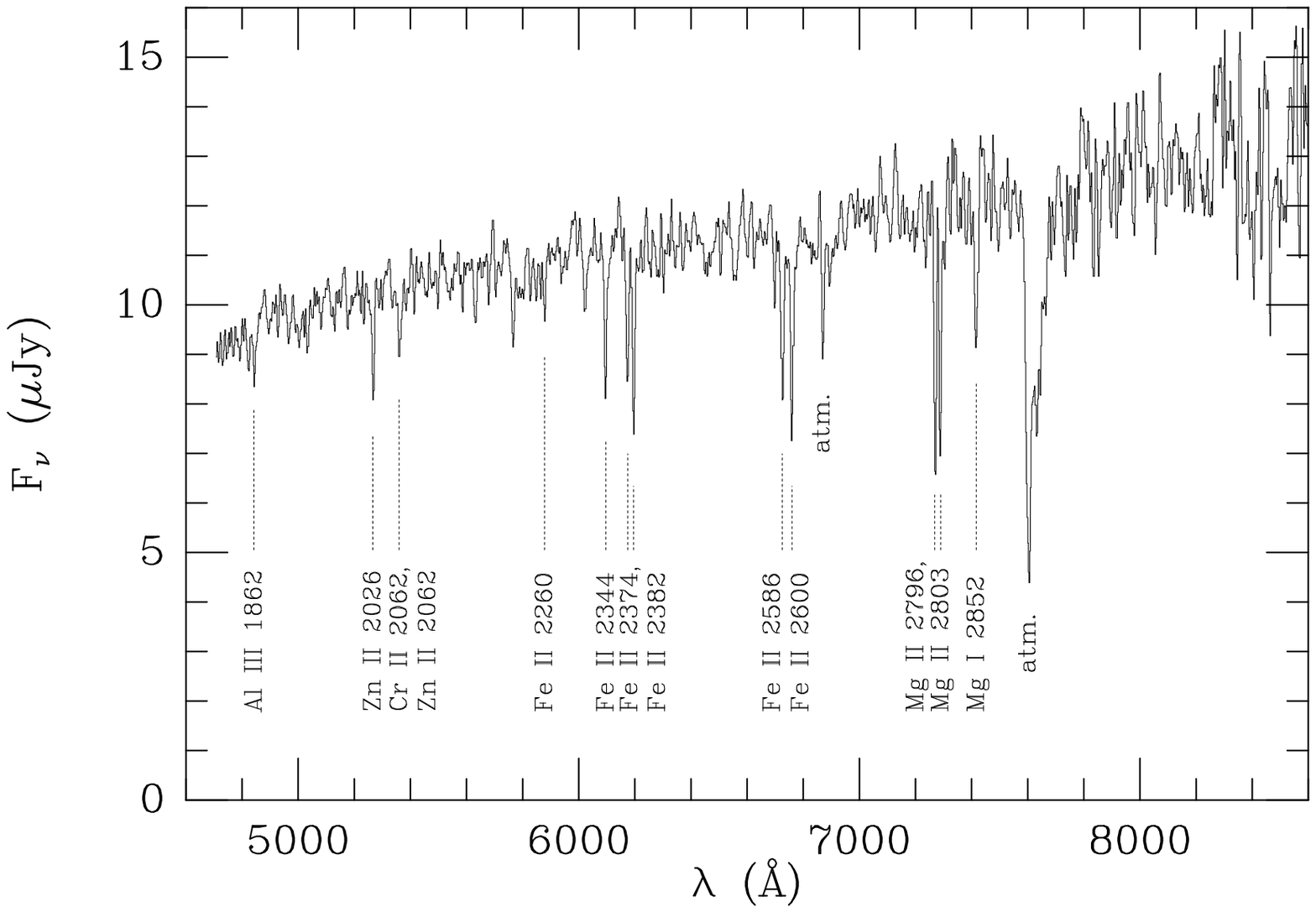}
\end{tabular}
\end{center}
\caption[example] { \label{fig:spectra} 
Examples of spectra of a grb host galaxy (GRB 970228, top), and afterglow
(GRB 990123, bottom), with the prominent emission and absorption features
labeled.  Both were obtained at the Keck telescope; see Bloom, Djorgovski,
\& Kulkarni (2001) and Kulkarni et al. (1999) for more details.
}
\end{figure} 

A new method for obtaining redshifts may come from the X-ray spectroscopy of
afterglows, using the Fe K line at $\sim 6.55$ keV (Piro et al. 1999, 2000;
Antonelli et al. 2000), or the Fe absorption edge at $\sim 9.28$ keV
(Weth et al. 2000; Yohshida et al. 1999; Amati et al. 2000).
Rapid X-ray spectroscopy of GRB afterglows may become a powerful
tool for understanding their physics and origins. 

\begin{figure}
\begin{center}
\begin{tabular}{c}
\bigskip\bigskip
\includegraphics[height=7cm]{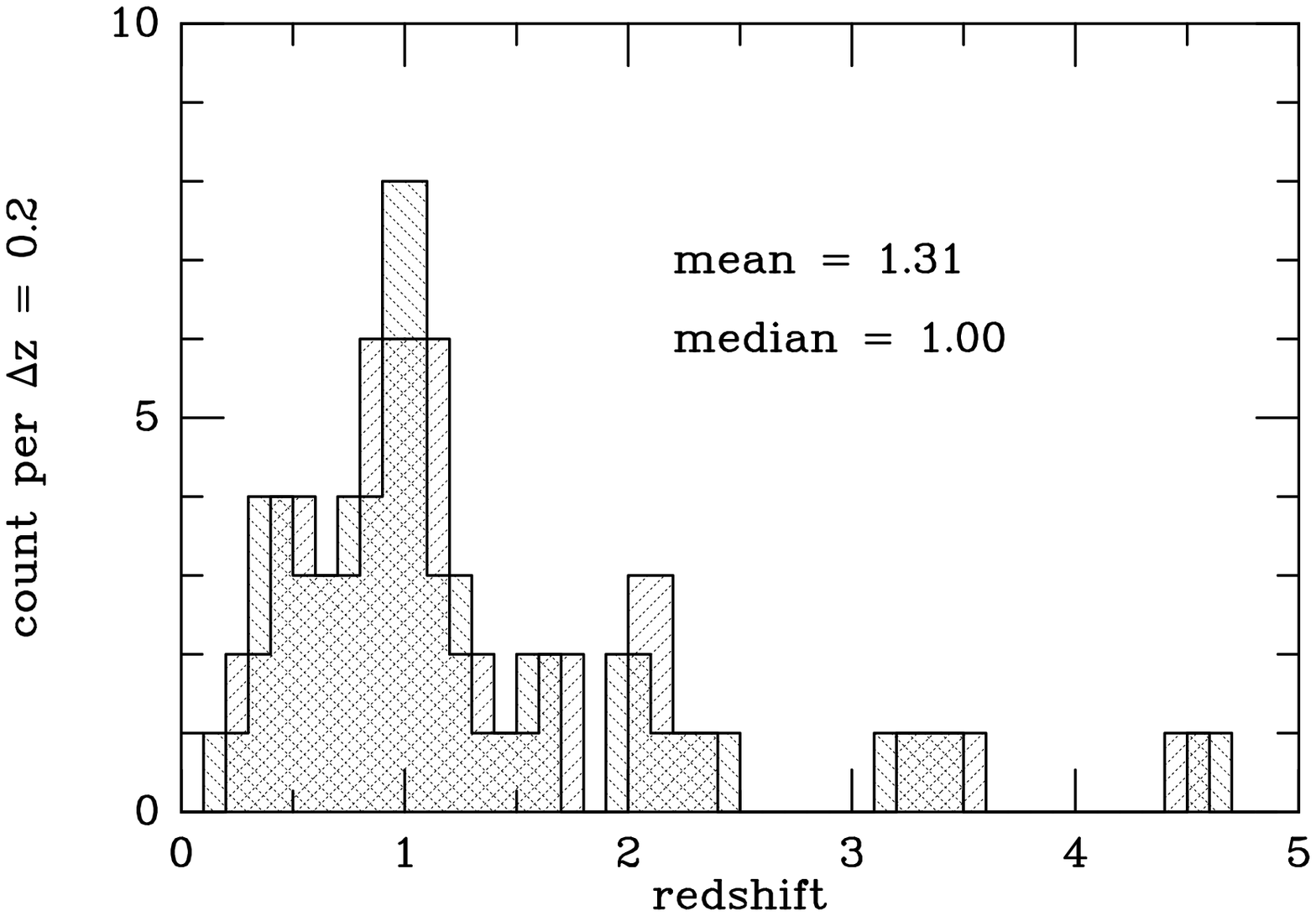}
\end{tabular}
\end{center}
\caption[example] { \label{fig:zhist} 
Redshift distribution of GRBs, as of November 2002.  The two hatched histograms
represent the same data, but with a binning offset by half a bin.
}
\end{figure} 

Are the GRB host galaxies special in some way?  If GRBs are somehow related to
massive star formation (e.g., Paczy\'nski 1998, Totani, 1997, etc.), it may be
worthwhile to examine their absolute luminosities and star formation rates
(SFR), or spectroscopic properties in general.  This is hard to answer
(Krumholz, Thorsett, \& Harrison 1998; Hogg \& Fruchter 1999; Schaefer 2000) 
from their visible ($\sim$ restframe UV) luminosities alone: the observed 
light traces an indeterminate mix of recently formed stars and an older
population, and cannot be unambiguously interpreted in terms of either the
total baryonic mass, or the instantaneous SFR.  

The magnitude and redshift distributions of GRB host galaxies are typical for
the normal, faint field galaxies, as are their morphologies
(Odewahn et al. 1998; Holland 2001; Bloom, Kulkarni \& Djorgovski 2002)
when observed with the HST: they are often compact, and sometimes suggestive
of a merging system (Djorgovski, Bloom \& Kulkarni 2001; Hjorth et al. 2002),
but that is not unusual for galaxies at comparable redshifts. 

Within the host galaxies, the distribution of GRB-host offsets follows 
the light distribution closely (Bloom, Kulkarni \& Djorgovski 2002), 
which is roughly proportional to the density of star formation (especially for
the high-$z$ galaxies).  It is thus fully consistent with a progenitor
population associated with the sites of massive star formation.

Spectroscopic measurements provide direct estimates of recent, massive
SFR in GRB hosts.  Most of them are based on 
the luminosity of the [O II] 3727 doublet (Kennicut 1998), 
the luminosity of the UV continuum at $\lambda_{rest} = 2800$ \AA\
(Madau, Pozzetti, \& Dickinson 1998),
in some cases (e.g., Kulkarni et al. 1998)
from the luminosity of the Ly$\alpha$ 1216 line,
and in others (e.g., Djorgovski et al. 1998)
from the luminosity of Balmer lines (Kennicut 1998).
All of these estimators are susceptible to the internal extinction and its
geometry, and have an intrinsic scatter of at least 30\%.
The observed $unobscured$ SFR's range from a few tenths to a few 
$M_\odot$ yr$^{-1}$.  Applying the reddening corrections derived from the
Balmer decrements of the hosts, or from the modeling of the broad-band colors
of the OTs (and further assuming that they are representative of the mean
extinction for the corresponding host galaxies) increases these numbers
typically by a factor of a few.  All this is entirely 
typical for the normal field galaxy population at comparable redshifts.
However, such measurements are completely insensitive to any fully obscured
SFR components.

Equivalent widths of the [O II] 3727 doublet in GRB hosts, which may provide
a crude measure of the SFR per unit luminosity (and a worse measure of the
SFR per unit mass), are on average somewhat higher (Djorgovski et al. 2001b)
than those observed in magnitude-limited field galaxy samples at comparable
redshifts (Hogg et al. 1998).
A larger sample of GRB hosts, and a good comparison sample, matched both 
in redshift and magnitude range, are necessary before any solid conclusions can
be drawn from this apparent difference. 

One intriguing hint comes from the flux ratios of [Ne III] 3869 to
[O II] 3727 lines: they are on average a factor of 4 to 5 higher in GRB
hosts than in star forming galaxies at low redshifts (Djorgovski et al. 2001b). 
Strong [Ne III] 
requires photoionization by massive stars in hot H II regions, and may represent
indirect evidence linking GRBs with massive star formation.

The interpretation of the luminosities and observed star formation rates is
vastly complicated by the unknown amount and geometry of extinction.  The
observed quantities (in the visible) trace only the unobscured stellar
component, or the components seen through optically thin dust.  Any 
stellar and star formation components hidden by optically thick dust cannot
be estimated at all from these data, and require radio and sub-mm observations.

Both observational windows, the optical/NIR (rest-frame UV) and the sub-mm
(rest-frame FIR)suffer from some biases: the optical band is significantly
affected by dust obscuration, while the sub-mm and radio bands lack
sensitivity, and therefore uncover only the most prodigiously star-forming
galaxies.  As of late 2002, radio and/or sub-mm emission powered by
obscured star formation has been detected from 4 GRB hosts
(Berger, Kulkarni \& Frail 2001; Berger et al. 2002b; Frail et al. 2002).
The surveys to date are sensitive only
to the ultra-luminous ($L > 10^{12} L_\odot$) hosts, with SFR of several
hundred M$_\odot$ yr$^{-1}$.  Modulo the uncertainties posed by the small
number statistsics, the surveys indicate that about 20\% of GRB hosts are
objects of this type, where about 90\% of the total star formation takes
place in obscured regions.

Given the uncertainties of the geometry of optically thin and optically thick
dust, optical colors of GRB hosts cannot be used to make any
meaningful statements about their net star formation activity.  The broad-band
optical colors of GRB hosts are not distinguishable from those of normal
field galaxies at comparable magnitudes and redshifts
(Bloom, Djorgovski, \& Kulkarni 2001; Sokolov et al. 2001; 
Chary, Becklin, \& Armus 2002; Le Floc'h et al. 2003).
It is notable that the optical/NIR colors of GRB hosts detected in the 
sub-mm are much bluer than typical sub-mm selected galaxies, suggesting that
the GRB selection may be probing a previously unrecognised population of
dusty star-forming galaxies.

On the whole, the GRB hosts seem to be representative of the normal,
star-forming
field galaxy population at comparable redshifts, and so far there is no
evidence for any significant systematic differences between them.

\subsection{GRB Hosts in the Context of Galaxy Evolution}

The observed redshift distribution of GRB hosts is about what is expected for
an evolving, normal
field galaxy population at these magnitude levels.  There is an excellent
qualitative correspondence between the observations and simple galaxy evolution
models (Mao \& Mo 1998).  

If GRB's follow the luminous mass, then the expected
distribution would be approximated by the luminosity-weighted galaxy luminosity
function (GLF) for the appropriate redshifts.
The hosts span a wide range of luminosities, with a characteristic absolute
restframe B band magnitude $M_{B,*} \approx -20$ mag, approximately half a
magnitude fainter than in the GLF at $z \approx 0$,
but comensurate with the late-type (i.e., star forming disk) galaxy
population at $z \approx 0$ (Madgwick et al. 2002; Norberg et al. 2002).  
This is somewhat surprising, since one
expects that the evolutionary effects would make the GRB host  galaxies,
with a typical $z \sim 1$, brighter than their descendants today.  The GRB
host GLF also has a somewhat steeper tail than the composite GLF at 
$z \approx 0$, but again similar to that of the star-forming, late-type
galaxies.  This is in a broad agreement with the results of deep redshift
surveys which probe the evolution of field galaxy populations out to $z \sim 1$
(Lilly et al. 1995; Ellis 1997; Fried et al. 2001; Lin et al. 1999).

The interpretation of these results is complex: the observed light reflects an
unknown combination of the unobscured fraction of recent star formation
(especially in the high-$z$ galaxies, where we observe the restframe UV
continuum) and the stellar populations created up to that point.  Our 
understanding of the field galaxy evolution in the same redshift range as
probed by the GRB hosts is still largely incomplete.  Different selection
effects may be plaguing the field and the GRB host samples.  While much
remains to be done, it seems that GRB hosts provide a new, independent
check on the traditional studies of galaxy evolution at moderate and high
redshifts.

\section{GRBs and Cosmology}

While interesting on their own, GRBs are now rapidly becoming powerful tools to
study the high-redshift universe and galaxy evolution, thanks to their apparent
association with massive star formation, and their brilliant luminosities.
GRBs and their afterglows should be readly detectable at large redshifts
(Lamb \& Reichart 2000).

There are three basic ways of learning about the evolution of luminous matter
and gas in the universe.  First, a direct detection of sources (i.e., galaxies)
in emission, either in the UV/optical/NIR (the unobscured components), or
in the FIR/sub-mm/radio (the obscured component).  Second, the detection of
galaxies selected in absorption along the lines of sight to luminous background
sources, traditionally QSOs.  Third, diffuse extragalactic backgrounds, which
bypass all of the flux or surface brightness selection effects plaguing all
surveys of discrete sources found in emission, but at a price of losing the
redshift information, and the ability to discriminate between the luminosity
components powered by star formation and powered by AGN.  Studies of GRB hosts
and afterglows can contribute to all three of these methodological approaches,
bringing in new, independent constraints for models of galaxy evolution and
of the history of star formation in the universe.

\subsection{Dark Bursts: Probing the Obscured Star Formation History}

Already within months of the first detections of GRB afterglows, no OT's were
found associated with some well-localised bursts despite deep and rapid
searches; the prototype ``dark burst'' was GRB 970828 (Djorgovski et al. 2001a).
Perhaps the most likely explanation for the non-detections of OT's when
sufficiently deep and prompt searches are made is that they are obscured by
dust in their host galaxies.  This is an obvious culprit if indeed GRBs are
associated with massive star formation.

Support for this idea also comes from detections of RTs without OTs, including
GRB 970828, 990506, and possibly also 981226 (see Frail et al. 2000 and
Taylor et al. 2000).  Dust reddening has been detected directly in some OTs 
(e.g., Ramaprakash et al. 1998; Bloom et al. 1998; Djorgovski et al. 1998, etc.);
however, this only covers OTs seen through optically thin dust, and there
must be others, hidden by optically thick dust.
An especially dramatic case was the RT (Taylor et al. 1998) and 
IR transient (Larkin et al. 1998)
associated with GRB 980329 (Yost et al. 2002).
We thus know that at least some GRB OTs must be obscured by dust. 

The census of OT detections for well-localised bursts can thus provide a
completely new and independent estimate of the mean obscured star formation
fraction in the universe.  Recall that GRBs are now detected out to 
$z \sim 4.5$ and that there is no correlation of the observed fluence with 
the redshift (Djorgovski et al. 2002),
so that they are, at least to a first approximation, good probes of the star
formation over the observable universe.  

As of late 2002, there have been $\sim 70$ adequately deep and rapid
searches for OTs from well-localised GRBs. 
We define ``adequate searches'' as reaching at least to $R \sim 20$ mag within
less than a day from the burst, and/or to at least to $R \sim 23 - 24$ mag
within 2 or 3 days; this is a purely heuristic, operational definition,
and an intentionally liberal one.  
In just over a half of such searches, OTs were found.
Inevitably, some OTs may have been missed due to an intrinsically low flux,
an unusually rapid decline rate (Fynbo et al. 2001; Berger et al. 2002a), 
or very high redshifts (so that the brightness in the commonly used $BVR$ bands
would be affected by the intergalactic absorption).
Thus the $maximum$ fraction of all OTs (and therefore massive star formation)
hidden by the dust is $\sim 50$\%.

This is a remarkable result.  It broadly agrees with the estimates that there
is roughly an equal amount of energy in the diffuse optical and FIR backgrounds
(see, e.g., Madau 1999).  This is contrary to some claims in the literature
which suggest that the fraction of the obscured star formation was much higher
at high redshifts.  Recall also that the fractions of the obscured and
unobscured star formation in the local universe are comparable.  

There is one possible loophole in this argument: GRBs may be able to destroy
the dust in their immediate vicinity (up to $\sim 10$ pc?)
(Waxman \& Draine 2000; Galama \& Wijers 2000),
and if the rest of the optical path through their hosts ($\sim$ kpc scale?)
was dust-free, OTs would become visible.  Such a geometrical arrangement may
be unlikely in most cases, and our argument probably still applies.
A more careful treatment of the dust evaporation geometry is needed, but
it is probably safe to say that GRBs can provide a valuable new constraint
on the history of star formation in the universe.

\subsection{GRBs as Probes of the ISM in Evolving Galaxies}

Absorption spectroscopy of GRB afterglows is now becoming a powerful new
probe of the ISM in evolving galaxies, complementary to the traditional studies
of QSO absorption line systems.  The key point is that the GRBs almost by
definition (that is, if they are closely related to the sites of
ongoing or recent massive star formation, as the data seem to indicate)
probe the lines of sight to dense, central regions of their host galaxies
($\sim 1 - 10$ kpc scale).  On the other hand, the QSO absorption systems
are selected by the gas cross section, and favor large impact parameters
($\sim 10 - 100$ kpc scale), mostly probing the gaseous halos of field galaxies,
where the physical conditions are very different.

The growing body of data on GRB absorption systems shows exceptionally high
column densities of gas, when compared to the typical QSO absorption systems;
only the highest column density DLA systems (themselves ostensibly star-forming
disks or dwarfs) come close (Savaglio, Fall, \& Fiore 2002; Castro et al. 2002;
Mirabal et al. 2002).
This is completely consistent with the general picture described above.
(We are refering here to the highest redshift absorbers seen in the afterglow
spectra, which are presumably associated with the host galaxies themselves;
lower redshift, intervening absorbers are also frequently seen, and their
properties appear to be no different from those of the QSO absorbers.)

This opens the interesting prospect of using GRB absorbers as a new probe of
the chemical enrichment history in galaxies in a more direct fashion than
what is possible with the QSO absorbers, where there may be a very complex
dynamics of gas ejection, infall, and mixing at play.  

Properties of the GRB absorbers are presumably, but not necessarily (depending
on the unknown geometry of the gas along the line of sight) reflecting the ISM
of the circum-burst region.  Studies of their chemical composition do not yet
reveal any clear anomalies, or the degree of depletion of the dust, but the
samples in hand are still too small to be really conclusive.  Also, there
have been a few searches for the variability of the column density of the gas
on scales of hours to days after the burst, with no clear detections so far.
Such an effect may be expected if the burst afterglow modifies the physical
state of the gas and dust along the line of sight by the evaporation of the
dust grains, additional photoionization of the gas, etc.  However, it is
possible that all such changes are observable only on very short time scales,
seconds to minutes after the burst.  In any case, a clear detection of a
variable ISM absorption against a GRB afterglow would be a very significant
result, providing new insight into the cisrcumstances of GRB origins.

\subsection{High-Redshift GRBs: Probing the Primordial Star Formation and Reionization}

Possibly the most interesting use of GRBs in cosmology is as probes of the
early phases of star and galaxy formation, and the resulting reionization of
the universe at $z \sim 6 - 20$.  If GRBs reflect deaths of massive stars,
their very existence and statistics would provide a superb probe of the
primordial massive star formation and the initial mass function (IMF).  
They would be by far the most 
luminous sources in existence at such redshifts (much brighter than SNe, and
most AGN), and they may exist at redshifts where there were $no$ luminous
AGN.  As such, they would provide unique new insights into the physics and
evolution of the primordial IGM during the reionization era (see, e.g.,
Lamb \& Reichart 2001; Loeb 2002a,b).

There are two lines of argument in support of the existence of copious numbers
of GRBs at $z > 5$ or even 10.  First, a number of studies using photometric
redshift indicators for GRBs suggests that a substantial fraction (ranging
from $\sim 10$\% to $\sim 50$\%) of all bursts detectable by past,
current, or forthcoming missions may be originating at such high redshifts,
even after folding in the appropriate spacecraft/instrument selection
functions (Fenimore \& Ramirez-Ruiz 2002; Reichart et al. 2001;
Lloyd-Ronning, Fryer, \& Ramirez-Ruiz 2002).

Second, a number of modern theoretical studies suggest that the very first
generation of stars, formed through hydrogen cooling alone, were very
massive, with $M \sim 100 - 1000 ~M_\odot$ (Bromm, Coppi \& Larson 1999;
Abel, Bryan, \& Norman 2000; Bromm, Kudritzki, \& Loeb 2001; Bromm, Coppi \& Larson 2002;
Abel, Bryan \& Norman 2002).
While it is not yet absolutely clear that some as-yet unforseen effect would
lead to a substantial fragmentation of a protostellar object of such a mass, a
top-heavy primordial IMF is at least plausible.  It is also not yet 
completely clear that the (probably spectacular) end of such an object
would generate a GRB, but that too is at least plausible (Fryer, Woosley \& Heger 2001).
Thus, there is some real hope that significant numbers of GRBs and their
afterglows would be detectable in the redshift range $z \sim 5 - 20$,
spanning the era of the first star formation and cosmic reionization (Bromm \& Loeb 2002).

Spectroscopy of GRB aftergows at such redshifts would provide a crucial,
unique information about the physical state and evolution of the primordial
ISM during the reionization era.  The end stages of the cosmic reionization
have been detected by spectroscopy of QSOs at $z \sim 6$
(Djorgovski et al. 2001c; Fan et al. 2001; Becker et al. 2001).
GRBs are more useful in this context than the QSOs, for several reasons.
First, they may exist at high redshifts where there were no comparably
luminous AGN yet.  Second, their spectra are highly predictable power-laws,
without complications caused by the broad Ly$\alpha$ lines of QSOs, and can
reliably be extrapolated blueward of the Ly$\alpha$ line.  Finally, they would
provide a genuine snapshot of the intervening ISM, without an appreciable
proximity effect which would inevitably complicate the interpretation of
any high-$z$ QSO spectrum (luminous QSOs excavate their Stromgren spheres
in the surrounding neutral ISM out to radii of at least a few Mpc, whereas
the primordial GRB hosts would have a negligible effect of that type;
see, e.g., Lazzati et al.(2001).

Detection of high-$z$ GRBs is thus an urgent cosmological task.  It requires
a rapid search for afterglows, as well as high-resolution
follow-up spectroscopy, in both the optical and NIR.  However, such effort would
be well worth the considerable scientific rewards in the end.

\acknowledgments     

We wish to thank numerous collaborators, and the staff of Palomar and
W.M. Keck Observatories
for their expert help during our observing runs.
Our work was supported by grants from the NSF, NASA, and private donors.


\begin{thebibliography}{1}


\bibitem{abn00}
  Abel, T., Bryan, G., \& Norman, M. 2000, ApJ, 540, 39

\bibitem{abn02}
  Abel, T., Bryan, G., \& Norman, M. 2002,  Science, 295, 93

\bibitem{apv+00}
  Antonelli, L.A., et al. 2000,  ApJ, 545, L39

\bibitem{beck+01}
  Becker, R., et al. (the SDSS collaboration) 2001,  AJ, 122, 2850
 
\bibitem{berger01}
  Berger, E., Kulkarni, S.R., \& Frail, D.A.  2001,  ApJ, 560, 652
  
\bibitem{berg+02}
  Berger, E., et al. 2002a, submitted to ApJ [astro-ph/0207320]
  
\bibitem{berger02}
  Berger, E., et al. 2002b, submitted to ApJ [astro-ph/0210645]

\bibitem{bfk+98}
  Bloom, J.S., et al. 1998, ApJ, 508, L21

\bibitem{bdk01}
  Bloom, J.S., Djorgovski, S.G. \& Kulkarni, S.R. 2001, ApJ, 554, 678

\bibitem{bkd02}
  Bloom, J.S., Kulkarni, S.R., \& Djorgovski, S.G. 2002, AJ, 123, 1111
  
\bibitem{bcl99}
  Bromm, V., Coppi, P., \& Larson, R. 1999,  ApJ, 527, L5

\bibitem{bcl02}
  Bromm, V., Coppi, P., \& Larson, R. 2002,  ApJ, 564, 23
    
\bibitem{bkl01}
  Bromm, V., Kudritzki, R., \& Loeb, A. 2001,  ApJ, 552, 464

\bibitem{bl02}
  Bromm, V., \& Loeb, A. 2002,  ApJ, 575, 111

\bibitem{bromm}
  Bromm, V., \& Schaefer, B. 1999, ApJ, 520,  661

\bibitem{bcs99}
  Brunner, R., Connolly, A., \& Szalay, A. 1999,  ApJ, 516, 563

\bibitem{castro03}
  Castro, S., et al. 2003, ApJ, in press

\bibitem{cba02}
  Chary, R., Becklin, E., \& Armus, L. 2002, ApJ, 566, 229

\bibitem{cos+97} Costa, E., et al. 1997, Nature, 387, 783 

\bibitem{dkb+98}
  Djorgovski, S.G., et al. 1998,  ApJ, 508, L17

\bibitem{sgd+01b}
  Djorgovski, S.G., et al. 2001a, ApJ, 562, 654

\bibitem{djorg01}
  Djorgovski, S.G., et al. 2001b, in Gamma-Ray Bursts in the Afterglow Era: 2nd Workshop,
  eds. E. Costa et al., ESO Astrophysics Symposia,  Berlin: Springer Verlag, p. 218 

\bibitem{dcsm01}
  Djorgovski, S.G., et al. 2001c, ApJ, 560, L5
  
\bibitem{sgd+01a}
  Djorgovski, S.G., Bloom, J.S., \& Kulkarni, S.R. 2002,  ApJ, in press [astro-ph/0008029]

\bibitem{mg9}
  Djorgovski, S.G., et al. 2002, in Proc. IX Marcel Grossmann Meeting,  eds. V. Gurzadyan et al. 
  Singapore: World Scientific, in press [astro-ph/0106574]

\bibitem{ellis97}
 Ellis, R. 1997,  ARAA, 35, 389

\bibitem{fan+01}
  Fan, X., et al. (the SDSS collaboration) 2001, AJ, 122, 2833
  
\bibitem{frr02}
  Fenimore, E., \& Ramirez-Ruiz, E. 2002, ApJ, in press [astro-ph/0004176]

\bibitem{fbg+00}
  Frail, D.A., et al. 2000,  ApJ, 538, L129

\bibitem{frail02}
  Frail, D.A., et al. 2002, ApJ, 565, 829

\bibitem{fr+01}
  Fried, J., et al. 2001,  A\&A, 367, 788

\bibitem{fwh01}
  Fryer, C., Woosley, S., \& Heger, A. 2001,  ApJ, 550, 372

\bibitem{fyn+01}
  Fynbo, J., et al. 2001,  A\&A, 369, 373
   
\bibitem{gw00}
  Galama, T.J., \& Wijers, R. 2000,  ApJ, 549, L209

\bibitem{hjorth2002}
  Hjorth, J., et al. 2002, ApJ, 576,  113

\bibitem{hol01}
  Holland, S. 2001,  [astro-ph/0102413]

\bibitem{hcbp98}
  Hogg, D., et al. 1998, ApJ, 504, 622

\bibitem{hf99}
  Hogg, D., \& Fruchter, A. 1999,  ApJ, 520, 54

\bibitem{hsd03}
  Hurley, K., Sari, R., \& Djorgovski, S.G. 2003, in 
  Compact Stellar X-Ray Sources, eds. W. Lewin \& M. van der Klis,
  Cambridge: Cambridge University Press 
  [astro-ph/0211620]

\bibitem{ken98}
  Kennicut, R. 1998, ARAA, 36, 131

\bibitem{kso73}
  Klebesadel, R., Strong, I., \& Olson, R. 1973, ApJ, 182, L85

\bibitem{kth98}
  Krumholtz, M., Thorsett, S., \& Harrison, F. 1998, ApJ, 506, L81

\bibitem{kul+98}
  Kulkarni, S.R., et al. 1998, Nature, 393, 35 
  
\bibitem{kul+99}
  Kulkarni, S.R., et al. 1999, Nature, 398, 389

\bibitem{kul+00}
  Kulkarni, S.R., et al. 2000, Proc. SPIE, 4005, 9

\bibitem{lamb00}
  Lamb, D., \& Reichart, D. 2000, ApJ, 536,  1

\bibitem{lr01}
  Lamb, D., \& Reichart, D. 2001,  in Gamma-Ray Bursts in the Afterglow Era: 2nd Workshop,
  eds. E. Costa et al., ESO Astrophysics Symposia,  Berlin: Springer Verlag, p. 226

\bibitem{lg+98}
  Larkin, J., et al. 1998,  GCN Circ. 44

\bibitem{lghfs01}
  Lazzati, D., et al. 2001,  in Gamma-Ray Bursts in the Afterglow Era: 2nd Workshop,
  eds. E. Costa et al., ESO Astrophysics Symposia, Berlin: Springer Verlag, p. 236

\bibitem{lefloch03}
  Le Floc'h, E., et al. 2003, A\&A, in press [astro-ph/0301149]

\bibitem{lth+95}
  Lilly, S., et al. 1995, ApJ, 455, 108

\bibitem{lin+99}
  Lin, H., et al. 1999,  ApJ, 518, 533
  
\bibitem{lrfrr02}
  Lloyd-Ronning, N., Fryer, C., \& Ramirez-Ruiz, E. 2002,  ApJ, 574, 554

\bibitem{loeb02a}
  Loeb, A. 2002a,  in Lighthouses of the Universe: The Most Luminous Celestial Objects and
  Their Use for Cosmology,  Eds. M. Gilfanov, R. Sunyaev \& E. Churazov, Berlin: Springer Verlag, p.137

\bibitem{loeb02b}
  Loeb, A. 2002b,  in Supernovae and Gamma-Ray Bursters, ed. K. Weiler,
  Berlin: Springer Verlag, in press  [astro-ph/0106455]

\bibitem{mpd98}
  Madau, P., Pozzetti, L.,\& Dickinson, M. 1998, ApJ, 498, 106

\bibitem{mad99}
  Madau, P. 1999,  ASPCS, 193, 475 

\bibitem{mad+02}
  Madgwick, D., et al. (the 2dF team) 2002, MNRAS, 333, 133
  
\bibitem{maomo98}
  Mao, S., \& Mo, H.J. 1998, A\&A, 339, L1
 
\bibitem{mesz01} 
  M\'esz\'aros, P. 2001, Science, 291, 79

\bibitem{mesz02} 
  M\'esz\'aros, P. 2002, ARAA, 40, 137
 
\bibitem{MR97} M\'esz\'aros, P., \& Rees M. J. 1997, ApJ, 476, 232

\bibitem{mr99} {M\'esz\'aros}, P. \& {Rees}, M. J. 1999, MNRAS, 306, L39

\bibitem{mdk+97} Metzger, M.R., et al. 1997, Nature 387, 879  

\bibitem{mirab02}
  Mirabal, M., et al. 2002, ApJ, 578, 818

\bibitem{nor+02}
  Norberg, P., et al. (the 2dF team) 2002, MNRAS, 336, 907

\bibitem{odk+98}
  Odewahn, S.C., et al. 1998, ApJ, 509, L5
 
\bibitem{pac98b}
  Paczy\'nski, B. 1998, ApJ, 494, L45

\bibitem{pian02}
  Pian, E. 2002, in Supernovae and Gamma-Ray Bursters, ed. K. W. Weiler, 
  Lecture Notes in Physics, Berlin: Springer-Verlag 
  [astro-ph/0110051]

\bibitem{piran99}
  Piran, T. 1999, Phys. Rep., 314, 575

\bibitem{pcf+99}
  Piro, L., et al. 1999, A\&ASup, 138, 431

\bibitem{pgg+00}
  Piro, L., et al. 2000,  Science, 290, 955

\bibitem{rkf+98}
  Ramaprakash, A., et al. 1998, Nature, 393, 43

\bibitem{reich+01}
  Reichart, D., et al. 2001,  ApJ, 552, 57

\bibitem{sff02}
  Savaglio, S., Fall, S.M., \& Fiore, F. 2002,  ApJ, in press

\bibitem{sch00}
  Schaefer, B. 2000,  ApJ, 532, L21

\bibitem{sfct+01}
 Sokolov, V.V., et al. 2001,  A\&A, 372, 428

\bibitem{tfk+98}
  Taylor, G.B., et al. 1998, ApJ, 502, L115

\bibitem{tbf+00}
  Taylor, G.B., et al. 2000, ApJ, 537, L17

\bibitem{tot97}
  Totani, T. 1997,  ApJ, 486, L71

\bibitem{pgg+97} 
  van Paradijs, J., et al. 1997, Nature, 386, 686 

\bibitem{vpkw00} 
  van Paradijs, J., Kouveliotou, C., \& Wijers, R. 2000, ARAA, 38, 379

\bibitem{wd00}
  Waxman, E., \& Draine, B. 2000, ApJ, 537, 796

\bibitem{wmkr00}
  Weth, C., et al. 2000, ApJ, 534, 581

\bibitem{yost+02}
  Yost, S., et al. 2002, ApJ, 577, 155
  
\end{thebibliography}
\end{document}